\documentclass[prl,twocolumn]{revtex4}
\usepackage{graphicx}
\usepackage{amssymb}
\usepackage{color}
\usepackage{enumerate}
\usepackage{amsmath}
\usepackage{graphics}
\newcommand{\be}{\begin{equation}}
\newcommand{\ee}{\end{equation}}
\newcommand{\bea}{\begin{eqnarray}}
\newcommand{\eea}{\end{eqnarray}}

\begin{document}

\title{Light polarization oscillations induced by photon-photon scattering}

\author{Fabio Briscese}\email{briscese.phys@gmail.com}

\affiliation{ Istituto Nazionale di Alta Matematica Francesco
Severi, Gruppo Nazionale di Fisica Matematica, Citt\`{a}
Universitaria, P.le A. Moro 5, 00185 Rome, Italy.}

\begin{abstract}

We consider the Heisenberg-Euler action for an electromagnetic
field in vacuum, which includes quantum corrections to the Maxwell
equations induced by photon-photon scattering. We show that, in
some configurations, the plane monochromatic waves become
unstable, due to the appearance of secularities in the dynamical
equations. These secularities can be treated using a multiscale
approach, introducing a slow time variable. The amplitudes of the
plane electromagnetic waves satisfy a system of ordinary
differential nonlinear equations in the slow time. The analysis of
this system shows that, due to the effect of photon-photon
scattering, in the unstable configurations the electromagnetic
waves oscillate periodically between left-hand-sided and
right-hand-sided polarizations. Finally, we discuss the physical
implications of this finding,  and the possibility of disclosing
traces of this effect in optical experiments.

\end{abstract}

\maketitle

\section{Introduction}\label{introduction}

Despite the fact that the equations of the classical
electromagnetic field in vacuum are linear, quantum corrections
due to photon-photon scattering introduce nonlinear effects.
Photon-photon scattering consists in the interaction of two
photons $\gamma_a$ and $\gamma_b$ of wave vectors $\vec{k}$ and
$\vec{h}$ that are scattered \textit{elastically}, so that after
the interaction the scattered photons $\gamma^\prime_a$ and
$\gamma^\prime_b$ will have wave vectors $\vec{k}^\prime$ and
$\vec{h}^\prime$. Of course, in order to scatter, the two
colliding photons   can not travel in the same direction, and this
trivial fact is reflected in the properties of the nonlinear terms
in the equations that we will study below.

The search for signatures of photon-photon scattering in optics is
still an open issue
\cite{lammerzal,jose,pike,Dinu:2014tsa,Dinu:2013gaa,king,di
piazza1,di piazza2,di piazza3,di piazza4,di piazza5,di piazza6,di
piazza7,di piazza8,di piazza9,di piazza10,di piazza11,di
piazza12,di piazza13,di piazza14,di piazza15,di piazza16,di
piazza17,di piazza18,di piazza19}, while indirect evidence for
this process  has been found in particle accelerators
\cite{denterria1,denterria2,denterria3,denterria4,denterria5,denterria6,denterria7,atlas}.

The quantum corrections due to photon-photon scattering were
calculated a long time ago by Heisenberg and Euler \cite{euler},
and extensively studied by other authors
\cite{dicus,Karplust,leo,schwinger}. The effective Lagrangian of
the electromagnetic field, obtained retaining only one electron
loop corrections, is \cite{schwinger}

\begin{equation}\label{lagrangian}
L = \frac{1}{4} F_{\mu\nu}F^{\mu\nu} + \epsilon^2 \left[ \left(
F_{\mu\nu}F^{\mu\nu}\right)^2 - \frac{7}{16} \left(
F_{\mu\nu}\tilde F^{\mu\nu}\right)^2 \right] \, ,
\end{equation}
where $F^{\mu\nu}= A^{\mu,\nu}-A^{\nu,\mu}$ is the electromagnetic
field \footnote{In this paper we use the covariant formalism, so
that the zeroth coordinate is defined as $x^0 = c \, t$.}, $A^\mu$
is the electromagnetic four-potential, $\tilde F^{\mu\nu} \equiv
\epsilon^{\mu\nu\alpha\beta}F_{\alpha\beta}$, and
\begin{equation}\label{epsilon}
\epsilon^2 = \frac{\alpha^2 \left(\hbar/m_e c \right)^3}{90 m_e
c^2} \, ,
\end{equation}
where $\alpha = e^2 /4\pi \epsilon_0 \hbar c\simeq 1/137$ is the
fine structure constant,  $\epsilon_0$  the dielectric
permeability of vacuum,  and $m_e$ the electron mass. Such a
Lagrangian is fit for this purpose when it is possible to neglect
other quantum effects. For instance, for low energetic photons of
energies $\textit{E}_\gamma   \ll m_e c^2$ particles creation is
inhibited, and the photon-photon scattering is the only process
inducing quantum correction to the Maxwell equations.

The terms $\propto \epsilon^2$ in the Lagrangian
(\ref{lagrangian}) take into account  photon-photon scattering,
and induce cubic corrections in the equations for the
four-potential $A^\mu$.    Since $\epsilon^2 \simeq 4\times
10^{-31} m^3/J$, so that $\epsilon^2 \, F_{\mu\nu}F^{\mu\nu}$ is
extremely small, such corrections are usually negligible with high
accuracy. However, in many physical situations, tiny perturbations
produce huge effects on a system, due to the action of hidden
resonances.

A paradigmatic example of this fact is given by rogue waves, which
are transient waves triggered by noise, appearing in several
physical contexts,   in particular in hydrodynamics
\cite{onorato}, nonlinear optics \cite{del re}, and Bose-Einstein
condensates \cite{bludov}. The simplest nonlinear model for the
description of such a phenomenon is the integrable focusing
Nonlinear Schr$\ddot{o}$dinger Equation (NLS) $i
\partial_t \psi + \partial^2_x \psi + 2 |\psi|^2 \psi = 0$
\cite{nls}, with the nonlinear mechanism of the modulation
instability \cite{modulation instability} at the basis of the
rogue waves generation (see also Refs. \cite{santini grinevich}
for recent theoretical developments).

In this paper we show that the dynamics  described by the
Lagrangian (\ref{lagrangian}) is unstable for some configurations
of the electromagnetic field. However, this instability is quite
different from the modulation instability \cite{modulation
instability} responsible for rogue waves generation, but it is not
less interesting. In fact, we find that, in the unstable
configurations, the electromagnetic waves oscillate periodically
between left-hand-sided and right-hand-sided polarizations, due to
the effect of photon-photon scattering.

We consider  two plane counterpropagating \footnote{As we will
see, the condition that the two waves are counterpropagating is
necessary in order to have non-vanishing nonlinear corrections.
This reflects the fact that such corrections comes from
photon-photon scattering, which require that the two photons have
opposite velocities to collide. } electromagnetic waves in vacuum,
and we show that the nonlinear terms in (\ref{lagrangian})
introduce resonant (or secular) corrections in the equations of
the electromagnetic field. Thus, in a standard perturbative
treatment, any small ($\sim \epsilon^2$) correction to the zeroth
order solution of the modified Maxwell equations will explode at a
finite time $T_i$, which is estimated as $T_i \sim 1/(\epsilon^2
\, c \, k\, <\rho>)$, where $k$ is the wavenumber of the
electromagnetic waves, and $<\rho>$ their energy-density.

Furthermore, we show that, introducing a slow  time variable, the
secular terms can be treated in a multiscale scheme.  In fact, the
amplitudes of the two counterpropagating waves satisfy a system of
nonlinear coupled ordinary differential equations in the slow
time. The analysis of this system shows that, for some initial
conditions, nonlinear corrections have negligible effects,
implying only small $\sim \epsilon^2$ corrections to the
energy-momentum dispersion relation of the photons; while for
other initial conditions, the effect of photon-photon scattering
is unexpectedly relevant, consisting of a continuous oscillation
in the polarization of the two beams between leaft-heand-sided and
right-hand-sided components.

Finally, we discuss the physical implications of this finding, and
we speculate on the possibility of detecting signatures of the
polarization oscillations in cosmological observations and in
optical experiments.

Without loss of generality, hereafter we will use the Lorentz
gauge
\begin{equation}\label{lorentz}
\partial_\alpha A^\alpha = 0 \, .
\end{equation}
Starting from the Lagrangian (\ref{lagrangian}) it is easy to show
that the modified Maxwell equations for the electromagnetic
four-potential $A^\alpha$ in the Lorentz gauge are

\begin{equation}\label{maxwell eq}
\begin{array}{ll}
\Box A^\alpha \left(1 + 8 \, \epsilon^2 \,
F_{\mu\nu}F^{\mu\nu}\right) +
\\
\\
+8 \, \epsilon^2\, \left[  F^{\alpha\beta} \partial_\beta
\left(F_{\mu\nu}F^{\mu\nu}\right) -  \frac{7}{16} \tilde
F^{\alpha\beta}
\partial_\beta \left( F_{\mu\nu}\tilde F^{\mu\nu}\right) \right] = 0 \,
,
\end{array}
\end{equation}
where $\Box$ is the d'Alembertian operator.

In principle, the smallness of the parameter $\epsilon$ justifies
a perturbative treatment of (\ref{maxwell eq}). However, as we
will se below, a naive perturbative approach is doomed to failure,
due to the presence of resonant terms.

At zeroth order in $\epsilon$,  Eq. (\ref{maxwell eq}) reduces to
the Maxwell equations in vacuum $\Box A^{\alpha} = 0$, which can
be solved exactly. Let us consider a zeroth order solution
$A^{(0)\alpha}$ corresponding to a system of two plane
electromagnetic waves propagating in the $x^3$ direction. Let us
express the four-potential $A^{(0)\alpha}$ in the form

\begin{equation}\label{plane waves order 0}
\left\{
\begin{array}{ll}
A^{(0)\alpha} = a^\alpha + b^\alpha + c.c. \, ,\\
\\
a^\alpha =  \xi^\alpha  \, e^{i k x} \, , \qquad b^\alpha =  \zeta^\alpha \,  e^{i h x} \\
\end{array}
\right.
\end{equation}
where c.c. stands for complex conjugate and the four-dimensional
polarization and wave vectors are constant and are given by
\begin{equation}\label{polarizzations plane waves order 0}
\left\{
\begin{array}{ll}
k=(k_0,0,0,k_3) \, , \qquad \xi=(0,\xi^1,\xi^2,0) \, ,\\
\\
h=(h_0,0,0,h_3) \, , \qquad \zeta=(0,\zeta^1,\zeta^2,0) \, ,
\end{array}
\right.
\end{equation}
with $|k_0/k_3|= |h_0/h_3|=1$.

We can study the evolution of the small perturbations of the
zeroth order solution (\ref{plane waves order 0}). Let us consider
a perturbative expansion of the four-potential in powers of
$\epsilon$ as

\begin{equation}\label{perturbative four vector}
A^\alpha = A^{(0)\alpha} + \epsilon^2 \delta A^{(2)\alpha} \,.
\end{equation}
Inserting this expression in (\ref{maxwell eq}) one has the
equations for the perturbations in the form

\begin{equation}\label{equations delta A}
\Box \delta A^{(2)\alpha}  + B^\alpha = 0 \, ,
\end{equation}
where we have defined
\begin{equation}\label{definition B}
\begin{array}{ll}
B^\alpha \equiv 8 \,  \left[  F^{\alpha\beta}
\partial_\beta \left(F_{\mu\nu}F^{\mu\nu}\right) -  \frac{7}{16}
\tilde F^{\alpha\beta}
\partial_\beta \left( F_{\mu\nu}\tilde F^{\mu\nu}\right) \right] \, .
\end{array}
\end{equation}
The four-vector $B^\alpha$ in (\ref{definition B}) must be
evaluated on the solution (\ref{plane waves order 0}). Direct
calculation gives

\begin{equation}\label{FF}
\begin{array}{ll}
F_{\mu\nu}F^{\mu\nu} = 2 \left[
\left(A^1_{\,\,\,,3}\right)^2-\left(A^1_{\,\,\,
,0}\right)^2+\left(A^2_{\,\,\,,3}\right)^2+\left(A^2_{\,\,\,,0}\right)^2
\right]=\\
\\
= 4 \left(k_0 h_0 - k_3 h_3 \right)  \left[ \left(\vec{\xi} \cdot
\vec{\zeta}\right)  e^{i(k+h)x} - \left(\vec{\xi} \cdot \vec{\bar
\zeta} \right) e^{i(k-h)x} \right] +  \\
\\
+ \, c.c. \, ,
\end{array}
\end{equation}
and
\begin{equation}\label{FF*}
\begin{array}{ll}
F_{\mu\nu}\bar F^{\mu\nu} = 8 \left[ A_{2,0} \,\, A_{1,3} -
A_{1,0} \, \, A_{2,3} \right]=\\
\\ 8 \left(k_0 h_3 - k_3 h_0 \right)
 \hat e^3 \cdot \left[
\left(\vec{\xi} \wedge  \vec{\zeta}\right)  e^{i(k+h)x} -  \left(\vec{\xi} \wedge \vec{\bar \zeta}\right)  e^{i(k-h)x} \right] + \\
\\
+ \, c.c. \, ,
\end{array}
\end{equation}
where $\vec{\xi}= (\xi^1,\xi^2,0)$, $\vec{\zeta}=
(\zeta^1,\zeta^2,0)$, $\hat e^3\equiv (0,0,1)$, "$\wedge$" and
"$\cdot$" are the vector and scalar products respectively and
$\vec{\bar \zeta}$ is the complex conjugate of $\vec{\zeta}$.
After some algebra one has

\begin{subequations}\label{partial FF}
\begin{equation}\label{partial FF 0 3}
\begin{array}{ll} F^{0 \beta} \partial_\beta
\left(F_{\mu\nu}F^{\mu\nu}\right) =F^{3 \beta} \partial_\beta
\left(F_{\mu\nu}F^{\mu\nu}\right) =0 \, ,
\end{array}
\end{equation}
\begin{equation}\label{partial FF 1 2}
\begin{array}{ll}
F^{j \beta} \partial_\beta \left(F_{\mu\nu}F^{\mu\nu}\right) = 4
\left(k_0 h_0 - k_3 h_3 \right)^2 \times \\
\\
\left\{ \left[ \left(\vec{\xi} \cdot \vec{\bar\zeta}\right)
\zeta^j + \left(\vec{\xi} \cdot \vec{\zeta}\right) \bar\zeta^j
\right] e^{ikx} + \left[ \left(\vec{\zeta} \cdot
\vec{\bar\xi}\right) \xi^j + \left(\vec{\zeta} \cdot
\vec{\xi}\right) \bar\xi^j \right] e^{ihx}  \right\} + \\
\\
+ c.c. + N.R.T.
\end{array}
\end{equation}
\end{subequations}

and
\begin{subequations}\label{partial FF*}
\begin{equation}\label{partial FF* 0 3}
\begin{array}{ll}
\tilde F^{0 \beta} \partial_\beta \left(F_{\mu\nu}\tilde
F^{\mu\nu}\right) = \tilde F^{3 \beta} \partial_\beta
\left(F_{\mu\nu} \tilde F^{\mu\nu}\right) =0 \, ,
\end{array}
\end{equation}

\begin{equation}\label{partial FF* 1 2}
\begin{array}{ll}
\tilde F^{j \beta} \partial_\beta \left(F_{\mu\nu} \tilde
F^{\mu\nu}\right) = 16 \left(k_0 h_0 - k_3 h_3 \right)^2 \epsilon^{jr} \times \\
\\
\left\{ \left[ \hat e^3 \cdot \left(\vec{\xi} \wedge
\vec{\bar\zeta}\right) \zeta^r + \hat e^3 \cdot \left(\vec{\xi}
\wedge
\vec{\zeta}\right) \bar\zeta^r \right] e^{ikx} +\right.\\
\\
\left. + \left[ \hat e^3 \cdot \left(\vec{\zeta} \wedge
\vec{\bar\xi}\right) \xi^r + \hat e^3 \cdot \left(\vec{\zeta}
\wedge
\vec{\xi}\right) \bar\xi^r \right]  e^{ihx}  \right\} + \\
\\
+ c.c. + N.R.T.
\end{array}
\end{equation}
\end{subequations}
where hereafter $j=1,2$ and N.R.T stands for non resonant terms.

First, we note that from Eq.s (\ref{partial FF 0 3}) and
(\ref{partial FF* 0 3}) one has $B^0 = B^3 = 0$, and therefore
from (\ref{equations delta A}) it follows that the components
$\delta A^{(2)0}$ and $\delta A^{(2)3}$ are stable.

Furthermore, from Eq.s (\ref{partial FF 1 2}) and (\ref{partial
FF* 1 2}) it follows that, if $k_0/k_3 = - h_0/h_3$, i.e.  the two
plane waves $a^\alpha$ and $b^\alpha$ in (\ref{polarizzations
plane waves order 0}) are counterpropagating, the components $B^j$
contain terms $\sim e^{i k x}$ and $\sim e^{i h x}$ that are
resonant, since they are a solution of the wave equation. Due to
such resonant terms, the components $\delta A^j$  grow linearly
with time,

\begin{equation}\label{exploding perturbations}
\delta A^j \sim \epsilon^2 \, k^3 \, (A^{(0)})^3 \, x^0 \, e^{i k
x} \sim  \epsilon^2 \, t \, c \, k^3 \, (A^{(0)})^3  \, e^{i k x}
\, ,
\end{equation}
where we have assumed for simplicity that $k_0 \sim h_0\sim k$,
and $|\vec{\xi}| \sim |\vec{\zeta}| \sim A^{(0)}$. From
(\ref{exploding perturbations})  it is evident that the
perturbative expansion in Eq. (\ref{perturbative four vector})
fails when $A^{(0)} \sim \epsilon^2 \delta A$, which gives the
time scale of the secularity as

\begin{equation}\label{timescale instability}
T_i \sim \frac{1}{c \, k^3 \, \epsilon^2 \left(A^{(0)}\right)^2}
\sim \frac{1}{c \, k \, \epsilon^2 <\rho>} \, ,
\end{equation}
where we $<\rho> \sim (k \, A^{(0)})^2$ is the energy density of
the electromagnetic field.

We note that, if $k_0/k_3 = h_0/h_3$, i.e., the two plane waves
$a^\alpha$ and $b^\alpha$ in (\ref{polarizzations plane waves
order 0}) propagate in the same direction, the nonlinear terms in
Eq. (\ref{maxwell eq}) disappear. In fact, in this case the
four-vector $A^\mu$ is a wave propagating in the positive $x^3$
direction, i.e., $A^\mu= A^\mu(x^0-x^3)$, or in the negative $x^3$
direction, i.e., $A^\mu= A^\mu(x^0+x^3)$. From (\ref{FF}) and
(\ref{FF*}) we can immediately  see that in such a case one has
$F_{\mu\nu} F^{\mu\nu}= 0$ and $F_{\mu\nu}\tilde F^{\mu\nu}= 0$.
Thus, from (\ref{definition B}) it comes that all the components
$B^\mu$ are null, and the zeroth order solution
(\ref{polarizzations plane waves order 0}) is an exact solution of
Eq. (\ref{maxwell eq}). Therefore, when the two waves $a^\mu$ and
$b^\mu$ propagate in the same direction, the effect of quantum
corrections due to photon-photon scattering disappear. This
reflects the fact that, in order to scatter, two photons can not
propagate in the same direction.

At that point, we focus on the case of two counterpropagating
plane waves, in which photon-photon scattering plays an important
role.   Hereafter, we consider the case $k_0=k_3 >0$ and
$h_0=-h_3>0$. In this case the occurrence of resonant terms in
(\ref{equations delta A}) implies the linear divergence
(\ref{exploding perturbations}) of the perturbation $\delta
A^\mu$. In what follows, we show that the failure of the
perturbative approach described above is amenable of a multiscale
treatment. In fact, in perturbation theory, the presence of
secularities is often due to a wrong perturbative approach, in
problems in which the solutions depend simultaneously on widely
different scales. In such cases, the divergences can be eliminated
introducing suitable slow variables, i.e., dealing with a
multiscale expansion. This analysis will finally clarify the
physical meaning of the secularities described above.

We introduce suitable  slow variables as

\begin{equation}\label{multiscale variables}
y^0 \equiv \epsilon^2 x^0 \,, \quad y^1 \equiv \epsilon \left(x^0+
x^3 \right) \,, \quad y^2 \equiv \epsilon \left(x^0 - x^3 \right)
\, .
\end{equation}
As we will see, the variables $y^1$ and $y^2$ will play no role in
the multiscale equations, so that the slow-scale evolution of the
system will depend only on the slow time $y^0$. Using
(\ref{multiscale variables})  one has
\begin{equation}\label{multiscale derivatives}
\begin{array}{ll}
\partial_{x^0} \rightarrow \partial_{x^0} + \epsilon
\left(\partial_{y^1} + \partial_{y^2}\right)+ \epsilon^2 \partial_{y^0} \, , \\
\\
\partial_{x^3} \rightarrow \partial_{x^3} + \epsilon
\left(\partial_{y^1} - \partial_{y^2}\right) \, ,
\end{array}
\end{equation}
which finally gives the d'Alembertian in terms of the derivatives
with respect to slow and fast variables as

\begin{equation}\label{multiscale d alembertian}
\begin{array}{ll}
\Box \rightarrow \Box + 2 \epsilon \left[ \left(\partial_{y^1} +
\partial_{y^2}\right) \partial_{x^0}
- \left(\partial_{y^1} - \partial_{y^2}\right) \partial_{x^3}
 \right] +\\
\\
+ 2 \epsilon^2 \left[ \partial_{x^0}
\partial_{y^0} +  2 \partial_{y^1} \partial_{y^2} \right] +
o(\epsilon^3)\, ,
\end{array}
\end{equation}
where $\Box$ is the d'Alembertian with respect to the fast
variables $x^0$ and $x^3$.

The multiscale approach is useful when the dynamics evolves on
widely different   scales. In this case, the dependence of the
solutions is split into fast and slow variables. Therefore, to
find meaningful (on long times) approximated solutions of Eq.
(\ref{maxwell eq}), we assume that the polarization vectors $\xi$
and $\zeta$ in (\ref{polarizzations plane waves order 0}) are no
longer constant, but depend on the slow variables $y^0$, $y^1$,
$y^2$. Moreover, we express the polarization vectors in terms of
the left and right polarizations $\hat e_L = (1,i,0)/\sqrt{2}$ and
$\hat e_R = (1,-i,0)/\sqrt{2}$ as

\begin{equation}\label{multiscale polarization vectors}
\begin{array}{ll}
\xi^\mu = a_L \, \hat e^L + a_R \, \hat e^R  \\
\\
\zeta^\mu = b_L \, \hat e^L + b_R \, \hat e^R
\end{array}
\end{equation}
where the coefficients $a_L$, $a_R$, $b_L$ and $b_R$ are the
complex amplitudes of the different polarizations of the
counterpropagating plane waves $a^\mu$ and $b^\mu$. Such
amplitudes depend on the slow variables only as
\begin{equation}\label{multiscale ab}
\begin{array}{ll}
a_L = a_L(y^0,y^1,y^2) \, , \quad a_R = a_R(y^0,y^1,y^2)\\
\\
b_L = b_L(y^0,y^1,y^2) \, , \quad b_R = b_R(y^0,y^1,y^2)
\end{array}
\end{equation}

Therefore, we search the solutions of Eq.s (\ref{maxwell eq}) in
the form (\ref{perturbative four vector}) with the conditions
(\ref{multiscale polarization vectors})-(\ref{multiscale ab}).
Inserting (\ref{perturbative four vector}) in (\ref{maxwell eq})
and using (\ref{multiscale d alembertian}) we have

\begin{equation}\label{multiscale equations}
\begin{array}{ll}
\left[\Box + 2 \epsilon \left[ \left(\partial_{y^1} +
\partial_{y^2}\right) \partial_{x^0}
- \left(\partial_{y^1} - \partial_{y^2}\right) \partial_{x^3}
 \right] + \right.\\
\\
\left.+  2 \epsilon^2 \left[ \partial_{x^0}
\partial_{y^0} +  2 \partial_{y^1} \partial_{y^2} \right]  \right] A^{(0)\alpha} + \epsilon^2 \Box \delta A^\alpha + \epsilon^2
B^\alpha = 0
\end{array}
\end{equation}

Equation (\ref{multiscale equations}) is automatically satisfied
at zeroth order in $\epsilon$, while at the first order $\sim
\epsilon$ it implies that the functions $a_L$, $a_R$, $b_L$, and
$b_R$ depend on the slow variables as

\begin{equation}\label{multiscale equations first order}
\begin{array}{ll}
a_L = a_L(y^0, y^1) \, , \quad a_R = a_R(y^0, y^1) \, \\
\\
b_L = b_L(y^0, y^2) \, , \quad b_R = a_R(y^0, y^2)\, .
\end{array}
\end{equation}

To write the second order equations in a compact form, we define
the vectors
\begin{equation}\label{vectors}
\begin{array}{ll}
\vec{A}^{(0)}= (0,A^{(0)1},A^{(0)2},0)\\
\\
\delta\vec{A}^{(0)}= (0,\delta A^1,\delta A^2,0)\\
\\
\vec{B}= (0,B^1,B^2,0).
\end{array}
\end{equation}

Using the fact that $A^{(0)0}=A^{(0)3}=\delta A^0=\delta
A^3=B^0=B^3= 0$, so that the 0 and 3 components of
(\ref{multiscale equations}) are automatically satisfied, at
second order $\sim \epsilon^2$ Eq. (\ref{multiscale equations})
gives

\begin{equation}\label{multiscale equations A}
2 \left[ \partial_{x^0} \partial_{y^0} +  2 \partial_{y^1}
\partial_{y^2} \right]  \vec{A}^{(0)} + R.T.(\vec{B}) = 0
\end{equation}
and
\begin{equation}\label{multiscale equations B}
\Box \delta \vec{A} + N.R.T. (\vec B) = 0 \, ,
\end{equation}
where $N.R.T.(\vec{B})$ and $R.T.(\vec{B})$ are, respectively, the
non resonant and resonant terms in $\vec{B}$.

Indeed, the terms $\sim \epsilon^2 A^{(0)\alpha}$ are used to
cancel the secularities contained in (\ref{multiscale equations}).
On the other hand, $\delta A^\alpha$ depends only on the fast
variables $x^0$ and $x^3$, and it is used to cancel the non
resonant terms in $B^\alpha$, so it is a stable and small
perturbation of $A^{(0)\alpha}$, and we neglect it in the
following discussion.

Let us write (\ref{multiscale equations A}) in explicit form.
Using (\ref{plane waves order 0}), (\ref{multiscale polarization
vectors}), (\ref{multiscale equations first order}) and
(\ref{vectors}) we have

\begin{equation}\label{second order A}
\begin{array}{ll}
\left[ \partial_{x^0} \partial_{y^0} +  2 \partial_{y^1}
\partial_{y^2} \right]  \vec{A}^{(0)}  = \\
\\
\left[ \left( i k_0 \partial_{y^0} a_L \right) \hat e_L + \left( i
k_0 \partial_{y^0} a_R \right) \hat e_R \right]e^{i k x} + \\
\\
\left[ \left( i h_0 \partial_{y^0} b_L \right) \hat e_L + \left( i
h_0 \partial_{y^0} b_R \right) \hat e_R \right]e^{i h x} + c.c.
\end{array}
\end{equation}
and
\begin{equation}\label{second order B}
\begin{array}{ll}
\vec{B}  = 32 \, k_0^2 \, h_0^2 \times \\
\\
\left\{ \left[ \left( -3 \, a_L \left(|b_L|^2+|b_R|^2 \right) + 22
\, a_R
b_L \bar b_R \right) \hat e_L + \right.\right.\\
\\
\left.\left. \left( -3 \, a_R \left(|b_L|^2+|b_R|^2 \right) + 22
\,
a_L b_R \bar b_L \right) \hat e_R \right] e^{i k x} \right.\\
\\
\left[ \left( -3 \, b_L \left(|a_L|^2+|a_R|^2 \right) + 22 \, b_R
a_L \bar a_R \right) \hat e_L + \right.\\
\\
\left.\left. \left( -3 \, b_R \left(|a_L|^2+|a_R|^2 \right) + 22
\,  b_L a_R \bar a_L \right) \hat e_R \right] e^{i k x} \right\}+\\
\\
+ c.c. + N.R.T.
\end{array}
\end{equation}
Substituting these expressions in (\ref{multiscale equations A}),
we obtain  the dynamical equations for the complex amplitudes as

\begin{equation}\label{multiscale equations amplitudes}
\begin{array}{ll}
i \partial_{y^0}a_L + 16 k_0 h_0^2 \left( -3 \, a_L
\left(|b_L|^2+|b_R|^2 \right) + 22 \, a_R
b_L \bar b_R \right)   = 0\\
\\
i \partial_{y^0}a_R + 16 k_0 h_0^2 \left( -3 \, a_R
\left(|b_L|^2+|b_R|^2 \right) + 22 \, a_L b_R \bar b_L \right)   =
0\\
\\
i \partial_{y^0}b_L + 16 k_0^2 h_0 \left( -3 \, b_L
\left(|a_L|^2+|a_R|^2 \right) + 22 \, b_R
a_L \bar a_R \right)   = 0\\
\\
i \partial_{y^0}b_R + 16 k_0^2 h_0 \left( -3 \, b_R
\left(|a_L|^2+|a_R|^2 \right) + 22 \, b_L a_R \bar a_L \right)   =
0 \, .
\end{array}
\end{equation}

Since the amplitudes $a_L$ and $a_R$ do not depend on $y^2$, while
$b_L$ and $b_R$ do not depend on $y^1$, the only possible
dependence on the variables $y^1$ and $y^2$ is

\begin{equation}\label{multiscale equations first order 2}
\begin{array}{ll}
a_L = \alpha_L(y^0) e^{i q y^1} \, , \quad a_R = \alpha_R(y^0) e^{i q y^1} \,, \\
\\
b_L = \beta_L(y^0) e^{i p y^2} \, , \quad b_R = \beta_R(i y^0)e^{i
p y^2} \, ,
\end{array}
\end{equation}
with $q$ and $p$ real arbitrary numbers. However, this exponential
dependence on $y^1$ and $y^2$ corresponds simply to an
$\epsilon^2$ correction to the wave vectors $k$ and $h$; moreover
$q$ and $p$ do not appear in (\ref{multiscale equations
amplitudes}), so we set $q=s=0$.

Therefore, hereafter the complex amplitudes will depend only on
the slow time $y^0$, and (\ref{multiscale equations amplitudes})
becomes an ordinary differential system.

Let us study (\ref{multiscale equations amplitudes}) in detail. It
is quite immediate to recognize that the energy densities
$<\rho_a> = k_0^2 \left(|a_L|^2+|a_R|^2 \right)$ and $<\rho_b> =
h_0^2 \left(|b_L|^2+|b_R|^2 \right)$ are constant. Therefore, the
intensities of the two plane waves $a^\mu$ and $b^\mu$ are
conserved separately.   Furthermore, the spin conservation implies
that the quantity $S= k_0 \left(|a_L|^2-|a_R|^2 \right)+ h_0
\left(|b_L|^2-|b_R|^2 \right)$ is also constant. Exploiting these
relations,  the system (\ref{multiscale equations amplitudes}) can
be simplified and then integrated (see the Appendix). However, to
have a better understanding of the dynamics under study,  we
continue our discussion of (\ref{multiscale equations
amplitudes}).

First, we study stable configurations. Let us consider the choice
$a_R = s_1   \, a_L$ and   $b_R = s_2 \, b_L$, where $s_1^2 =
s_2^2 =1$ for  two counterpropagating linearly polarized beams,
and $s_1 = s_2 = 0$ for  two left-handed circularly polarized
waves (different choices, e.g., $a_L = s_1 a_R$ and $b_L = s_2
b_R$, give identical results).  In that case, (\ref{multiscale
equations amplitudes}) becomes

\begin{equation}\label{multiscale equations amplitudes stable 2}
\begin{array}{ll}
i a^\prime_L +16 k_0 h_0^2 \left[-3\left(1+s_2^2 \right) + 22 \, s_1 \, s_2 \right] a_L  |b_L|^2   =0\\
\\
i b^\prime_L +16 k_0^2 h_0 \left[-3\left(1+s_1^2 \right) + 22 \,
s_1 \, s_2 \right] b_L |a_L|^2   =0 \, .
\end{array}
\end{equation}
and its solution is

\begin{equation}\label{multiscale equations amplitudes stable solutions 2}
a_R = s_1 \, a_L = s_1 \,  a^0_L e^{i \omega y^0}\, \quad b_R =
s_2 \,  b_L = s_2 \,  b^0_L e^{i \gamma y^0} \, ,
\end{equation}
where the frequencies $\omega$ and $\gamma$ are given by

\begin{equation}\label{multiscale equations amplitudes stable solutions frequency 2}
\begin{array}{ll}
\omega = 16\left[-3\left(1+s_2^2 \right) + 22 \, s_1 \, s_2 \right] k_0 h_0^2 |b^0_L|^2 \\
\\
\gamma = 16 \left[-3\left(1+s_1^2 \right) + 22 \, s_1 \, s_2
\right] k_0^2 h_0 |a^0_L|^2
\end{array}
\end{equation}

In this class of solutions the complex amplitudes have constant
modulus, and the quantum corrections  affect only the phase of the
waves. The only effect of the nonlinearity is a correction to the
frequency of light given by

\begin{equation}\label{multiscale equations amplitudes stable solutions corrections frequency}
k_0^\prime = k_0 + \epsilon^2 \omega  \, \quad h_0^\prime = h_0 +
\epsilon^2 \gamma \, ,
\end{equation}
and the  relative frequency shifts are

\begin{equation}\label{multiscale equations amplitudes stable solutions frequency shifts}
\begin{array}{ll}
\frac{\Delta k_0}{k_0} \sim \epsilon^2 h_0^2 |b^0_L|^2\sim
\epsilon^2 <\rho_b>  \\
\\
\frac{\Delta h_0}{h_0} \sim \epsilon^2 k_0^2 |a^0_L|^2\sim
\epsilon^2 <\rho_a> \, .
\end{array}
\end{equation}

We mention that many quantum gravity models predict a violation of
the Lorentz symmetry \cite{amelino}, implying a deformation of the
photon energy-dispersion relation, that might be observed in
photons of astrophysical origin, e.g., in gamma ray bursts
\cite{amelino2}. However, one should keep in mind that an apparent
breaking of the Lorentz symmetry in photons  might be due to
photon-photon scattering rather than quantum gravitational
effects.

\begin{figure}[tbp]
\begin{center}
\includegraphics[width=3.7in, scale=1]{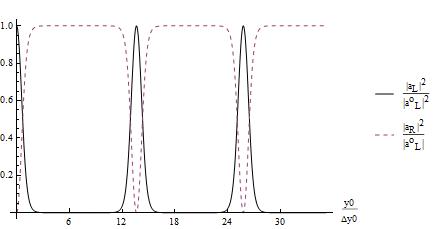} \caption{ We plot the
evolution  of  $|a_L|^2/|a^0_L|^2$ (solid line) and
$|a_R|^2/|a^0_L|^2$ (dashed line) against the slow time $y^0$ in
units of $\Delta Y^0$. The plot shows the oscillatory behavior of
the polarization of the light beam $a^\mu$. } \label{fig1}
\end{center}
\end{figure}

\begin{figure}[tbp]
\begin{center}
\includegraphics[width=3.7in, scale=1]{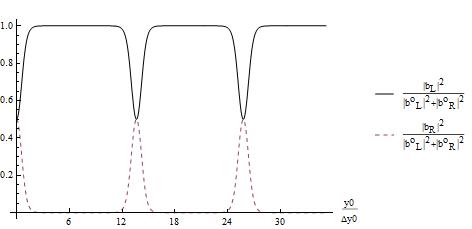} \caption{ We plot the
evolution  of  $|b_L|^2/(|b^0_L|^2+|b^0_R|^2)$ (solid line) and
$|b_R|^2/(|b^0_L|^2+|b^0_R|^2)$ (dashed line) against the slow
time $y^0$ in units of $\Delta Y^0$. The plot shows the
oscillatory behavior of the polarization of the light beam
$b^\mu$, with weaker  oscillations with respect to the beam
$a^\mu$. } \label{fig2}
\end{center}
\end{figure}

At that point, we study most interesting configurations, in which
the initial polarizations of the light beams change dramatically
during the evolution of the system. We choose initial conditions
in such a way that at least one of the products $a_L \, a_R$ or
$b_L \, b_R$ is initially nonzero. In facts, the last terms in Eq.
(\ref{multiscale equations amplitudes}) are responsible of the
oscillatory behavior that we describe below. Solving
(\ref{multiscale equations amplitudes}) numerically it is possible
to see that the polarization of the two counterpropagating waves
oscillate periodically between left-handed and right-handed
polarizations. For instance, in Fig.s \ref{fig1} and \ref{fig2} we
plot the square modulus of the amplitudes for $k_0=h_0=1$ and
initial values $a^0_R = 0$, $a^0_L = 1$, $b^0_L = 1$, $b^0_R = i$.
From (\ref{multiscale equations amplitudes}) it is easy to
recognize that the time scale of variation of the amplitudes is
given by (\ref{timescale instability}), thus we plot the solutions
in a time interval $\Delta t \sim 30 \, T_i$ corresponding to an
interval $\Delta y^0 = \epsilon^2 \Delta t$, in which the
oscillatory behavior of the dynamics is shown completely.

From Fig. \ref{fig1} we see that $|a_R|$ is initially zero, but it
grows rapidly  to $|a_R|=|a^0_L|$, while $|a_L|$ goes from
$|a^0_L|$ to zero. Thus, the $a^\mu$ beam initially in the
left-handed polarization   switches to the right-handed
polarization. It remains in this state most of the time, until it
jumps back to its initial left-handed configuration after the
first period  $\tau \sim 13 \, T_i$. After that, this behavior is
repeated periodically.

The behavior of the beam $b^\mu$ is similar. In facts, $|b_R|$
goes to zero rapidly, while $|b_L|$ goes to
$\sqrt{|b^0_L|^2+|b^0_R|^2}$, and it remains constant most of the
time until, after a time  $\tau \sim 13 \, T_i$,   $|b_L|$ and
$|b_R|$ go back to their initial values. The difference with
respect the beam $a^\mu$ is that $|b_L|$ does never reach the
zero.

We emphasize that the reason why the system spends most of the
time in the configuration $|a_L| \simeq |b_R| \simeq 0$, $|a_R|
\simeq |a^0_L|$, and $|b_L| \simeq \sqrt{|b^0_L|^2+|b^0_R|^2}$ is
that this configuration corresponds approximately to  two
circularly polarized counterpropagating beams, which is analogous
to the solution (\ref{multiscale equations amplitudes stable
solutions 2}), which is stable.

Numerical investigation of (\ref{multiscale equations amplitudes})
suggests that this oscillatory behavior is not affected
(qualitatively) by the choice of the parameters in
(\ref{multiscale equations amplitudes}), while the recurrence time
of the polarization oscillations depends on the wavelength and
energy density of the electromagnetic field as in (\ref{timescale
instability}).

Finally, we note that the equivalence between the space and time
coordinates $x^0$ and $x^3$, which is evident in the covariant
formalism, implies that the oscillatory behavior may occur also in
the space variable $x^3$. However, this issue goes beyond the
purpose of this paper and will be discussed elsewhere
\cite{Briscese new}.

To finish, we estimate  the recurrence times for realistic
physical situations in cosmology and optics.

Due to the extreme smallness of $\epsilon^2$, one expects a huge
recurrence time $T_i$. For that reason, it is natural expect a
$T_i$  of the order of cosmological times $1/H_0 \sim 10^{11} s$
\footnote{$H_0$ is the hubble constant today, and its inverse
represents a typical cosmological timescale.}. Indeed, we can ask
whether this instability might be important in cosmology, so we
can estimate $T_i$ in the case of the cosmic microwave background
(CMB) radiation \cite{planck}, which is an almost perfect
blackbody radiation with a temperature of $ 2.7 \, K$. The energy
density of the CMB radiation is measured as $<\rho> \sim 10^{-14}
J/m^3$, while its wavelength is in the microwave range $\lambda
\sim 1 mm$, which gives a time   $T_i \simeq 10^{42} \, s$  much
greater than the age of the universe. It might be argued that it
could be possible to have $T_i \sim 1/H_0$ in the early universe.
However, at high redshifts photons are highly energetic, and
therefore they can produce other particles (e.g.,
electron-positron pairs). Indeed, other quantum effects are not
negligible, so that the Lagrangian (\ref{lagrangian}) is no longer
fit for our purposes.  Therefore, one concludes that the
polarization oscillations can not be observed in the CMB
radiation.

Finally, we consider the possibility of observing the polarization
oscillations in optical experiments. The search for signatures of
the photon-photon scattering in optics is in progress
\cite{lammerzal,jose,pike,Dinu:2014tsa,Dinu:2013gaa,king,di
piazza1,di piazza2,di piazza3,di piazza4,di piazza5,di piazza6,di
piazza7,di piazza8,di piazza9,di piazza10,di piazza11,di
piazza12,di piazza13,di piazza14,di piazza15,di piazza16,di
piazza17,di piazza18,di piazza19}. For instance, Ref.  \cite{di
piazza1}  investigates the possibility of observing vacuum
birefringence and dichroism induced by photon-photon interactions
in ultra-strong laser fields.

We can estimate the time of recurrence of the polarization
oscillations for light beams produced in petawatt class lasers,
which will be available in the near future. The intensities
attainable in these lasers reach $I \sim 10^{23} W/cm^2$
\cite{lasers1,lasers2}, giving a recurrence time  $T_i \sim 4
\times 10^2 \left(\lambda/m \right) \, s$, where $\lambda/m$ is
the laser wavelength in meters (we used $k \sim h \sim 2
\pi/\lambda$ and $k^2 a^2 \sim k^2 b^2 \sim <\rho> \sim I/c$).
Therefore, for realistic lasers  with $\lambda \sim 1 \, \mu m$,
observation times can be of the order of $10^{-3 } \, s$ (to be
compared with those estimated in Ref. \cite{di piazza1}). This
lets us hope to be able to observe polarization oscillations in
two counterpropagating petawatt laser beams.

In conclusion, in this paper we have shown that the extremely weak
photon-photon interaction might be responsible for surprisingly
strong deviations from the free dynamics of electromagnetic waves.
We have shown that, in the case of two counterpropagating laser
beams, one of which has circular polarization and the other is not
circularly  polarized, the evolution of the electromagnetic waves
consists in slow oscillations in the polarizations of the beams.
We have estimated the recurrence of the polarizzation
oscillations, and we have shown that, while unobservable in the
cosmological context, this oscillatory behavior might be revealed
in realistic optical experiments.

\textbf{Acknowledgments:} The author is very grateful to P.
Santini, F. Calogero and E. Del Re for useful discussions on the
draft version of this paper.

\appendix

\section{Appendix} \label{appendix}

In this appendix we show that it is possible to  solve the system
(\ref{multiscale equations amplitudes}). First, it is immediate to
recognize that (\ref{multiscale equations amplitudes}) implies the
conservations of the energy densities $<\rho_a> = k_0^2
\left(|a_L|^2+|a_R|^2 \right)$ and $<\rho_b> = h_0^2
\left(|b_L|^2+|b_R|^2 \right)$. Therefore, it is natural to
express the amplitudes in the form

\begin{equation}\label{amplitudes reduced}
\begin{array}{ll}
a_L(y^0) = a^0 \, \sin\left(\phi\left(y^0\right)\right) e^{i \left(\omega y^0 + \theta_L^0\right)}\\
\\
a_R(y^0) = a^0 \cos\left(\phi\left(y^0\right)\right) e^{i \left(\omega y^0 + \theta_R^0\right)}\\
\\
b_L(y^0) = b^0 \, \sin\left(\varphi\left(y^0\right)\right) e^{i \left(\gamma y^0+ \psi_L^0\right)}\\
\\
b_R(y^0) = b^0 \cos\left(\varphi\left(y^0\right)\right) e^{i\left(
\gamma y^0 + \psi_R^0\right)} \, ,
\end{array}
\end{equation}
where   $a^0$, $b^0$, $\theta_L^0$, $\theta_R^0$, $\psi_R^0$ and $
\psi_L^0$ are arbitrary constants such that $\theta_R^0-\theta_L^0
+\psi_L^0-\psi_R^0 = \pi/2 $ and
\begin{equation}\label{frequencies reduced}
\omega = -48 \, k_0 \, h_0^2 \, |b^0|^2 ,\qquad \gamma = -48 \,
k_0^2 \, h_0 \, |a^0|^2 \, .
\end{equation}

Substituting (\ref{amplitudes reduced}) in (\ref{multiscale
equations amplitudes}), one obtains  the following reduced system
for the two variables $\phi$ and $\varphi$:

\begin{equation}\label{reduced system}
\begin{array}{ll}
\phi^\prime + 176 \, k_0 h_0^2 |b^0|^2 \sin\left(2\varphi \right)= 0 \\
\\
\varphi^\prime - 176 \, k_0^2 h_0 |a^0|^2 \sin\left(2\phi \right)
= 0 \, ,
\end{array}
\end{equation}
that can be solved using the spin conservation $k_0 \, |a^0|^2
\cos\left(2\phi \right)+ h_0 \, |b^0|^2 \cos\left(2\varphi
\right)+ S= 0$. Therefore (\ref{multiscale equations amplitudes})
is integrable and its solutions are periodic. However, we
preferred to discuss (\ref{multiscale equations amplitudes})
instead of (\ref{reduced system}), since  the separate analysis of
the behavior of the four amplitudes $a_L$, $a_R$, $b_L$, and $b_R$
makes   the oscillatory dynamics of the polarizations more
evident.

\end{document}